\def\year{2024}\relax
\documentclass[letterpaper]{article} 
\usepackage{aaai24}  
\usepackage{times}  
\usepackage{helvet}  
\usepackage{courier}  
\usepackage[hyphens]{url}  
\usepackage{graphicx} 
\usepackage{subcaption}
\usepackage{natbib}  

\usepackage{caption} 
\DeclareCaptionStyle{ruled}{labelfont=normalfont,labelsep=colon,strut=off} 
\usepackage{algorithm}
\usepackage{algorithmic}
\usepackage{xspace}
\usepackage[most]{tcolorbox}

\newcommand{\one}{({\em i}\/)\xspace}
\newcommand{\two}{({\em ii}\/)\xspace}
\newcommand{\three}{({\em iii}\/)\xspace}
\newcommand{\four}{({\em iv}\/)\xspace}
\newcommand{\five}{({\em v}\/)\xspace}
\newcommand{\six}{({\em vi}\/)\xspace}

\def\eg{\emph{e.g.}\xspace}
\def\ie{\emph{i.e.}\xspace}

\newcommand{\pb}[1]{\vspace{0.75ex}\noindent{\bf \em #1}\hspace*{.3em}}

\usepackage{xcolor}
\newcommand{\answerYes}[1]{\textcolor{blue}{#1}} 
 
\newcommand{\answerNA}[1]{\textcolor{gray}{#1}}

\usepackage{newfloat}
\usepackage{listings}
\floatstyle{ruled}
\newfloat{listing}{tb}{lst}{}
\floatname{listing}{Listing}

\title{Decentralised Moderation for Interoperable Social Networks: 
\\ A Conversation-based Approach for Pleroma and the Fediverse}

\author{
    Vibhor Agarwal\textsuperscript{\rm 1}, 
    Aravindh Raman\textsuperscript{\rm 2},
    Nishanth Sastry\textsuperscript{\rm 1}, \\ 
    Ahmed M. Abdelmoniem\textsuperscript{\rm 3},
    Gareth Tyson\textsuperscript{\rm 4},
    Ignacio Castro\textsuperscript{\rm 3}
}
\affiliations{
    \textsuperscript{\rm 1}University of Surrey, Guildford, United Kingdom \\
    \textsuperscript{\rm 2}Telefonica Research, Barcelona, Spain    \\
    \textsuperscript{\rm 3}Queen Mary University of London, London, United Kingdom    \\
    \textsuperscript{\rm 4}Hong Kong University of Science and Technology (GZ), China    \\

    \{v.agarwal, n.sastry\}@surrey.ac.uk, aravindh.raman@telefonica.com, \\ \{ahmed.sayed, i.castro\}@qmul.ac.uk, gtyson@ust.hk
}

\begin{document}

\maketitle

\begin{abstract}

The recent development of decentralised and interoperable social networks (such as the ``fediverse'') creates new challenges for content moderators. This is because millions of posts generated on one server can easily ``spread'' to another, even if the recipient server has very different moderation policies. 
An obvious solution would be to leverage moderation tools to automatically tag (and filter) posts that contravene moderation policies, \eg related to toxic speech.
Recent work has exploited the conversational context of a post to improve this automatic tagging, \eg using the replies to a post to help classify if it contains toxic speech.
This has shown particular potential in environments with large training sets that contain complete conversations.
This, however, creates challenges in a decentralised context, as a single conversation may be fragmented across multiple servers. 
Thus, each server only has a partial view of an entire conversation because conversations are often federated across servers in a non-synchronized fashion.
To address this, we propose a decentralised conversation-aware content moderation approach suitable for the fediverse. Our approach employs a graph deep learning model (GraphNLI) trained locally on each server. The model exploits local data to train a model that combines post and conversational information captured through random walks to detect toxicity.
We evaluate our approach with data from Pleroma, a major decentralised and interoperable micro-blogging network containing $2$ million conversations.
Our model effectively detects toxicity on larger instances, exclusively trained using their local post information ($0.8837$ macro-F1).
Yet, we show that this approach does not perform well on smaller instances that do not possess sufficient local training data.
Thus, in cases where a server contains insufficient data, we strategically retrieve information (posts or model parameters) from other servers to reconstruct larger conversations and improve results. With this, we show that we can attain a macro-F1 of $0.8826$. Our approach has considerable scope to improve moderation in decentralised and interoperable social networks such as Pleroma or Mastodon.

\end{abstract}

\section{Introduction}
Network effects and economies of scale favour concentration~\cite{varian2019recent}.
As a result, the Web has witnessed increasing concentration into a few key players~\cite{doan2022empirical}.
For social networks, interoperability has been proposed as a remedy~\cite{beck2023breaking} and the Digital Markets Act (DMA) will soon enforce it in the EU~\cite{noauthor_q_nodate, ietf-mimi-content-00}.
We believe this opens new opportunities and challenges, particularly with regards to content moderation.

To understand these we look into the fediverse, an ecosystem of decentralised, interoperable platforms
powered by a common underlying protocol, ActivityPub~\cite{activitypub2018}.
In contrast to X (previously, Twitter), Reddit and other centralised social networks, in the fediverse anyone can create an instance of a service by running open-source code on a server.
Users can then sign up with such instances and use their services.
Fediverse users can interact with each other regardless of the
instance they have signed up with or even the service.
For instance, users from an instance in the micro-blogging Pleroma can interact with users from another Pleroma instance,
as well as with users from instances across other services such as Mastodon (micro-blogging) or Peertube (video streaming).
The (coming) support for ActivityPub by new services such as Threads,\footnote{\url{https://engineering.fb.com/2023/09/07/culture/threads-inside-story-metas-newest-social-app/}, last accessed 14 Sep 2023.} may expand this interoperable ecosystem even further.

Decentralisation and interoperability, however, pose new challenges.
Decentralisation limits the human and data resources required for (semi) automating moderation.
Interoperability limits the visibility and ability to affect content generated in a remote instance.
In contrast to other volunteer-moderated communities (\eg, Reddit)~\cite{birman2018moderation,iqbal2022exploring}, in the fediverse an administrator may need to handle content originating from a different instance or even service (\eg Mastodon posts frequently appear in Pleroma instances). 
With toxic content being common in the fediverse~\cite{hassan2021exploring, bin2022toxicity},
moderation can be challenging for administrators~\cite{anaobi2023will}.

We believe that this is a challenge that further interoperability (\eg, due to Threads supporting ActivityPub or thanks to DMA) will exacerbate.
To study this problem and identify solutions, 
we use a comprehensive dataset comprising $713$ instances and $16.5$ million posts from a large  micro-blogging fediverse service, Pleroma~\cite{bin2022toxicity}. 
Our main assumption is the usage of ActivityPub and, as such, our solutions should be amenable to other fediverse services (\eg, Mastodon).
Our goal is to validate the viability of our approach rather than necessarily advocate for Pleroma or other fediverse platform to adopt it.

To cope with the limitations in terms of the data that an instance has available, we take a conversational approach\footnote{The code is publicly available at \url{https://github.com/vibhor98/decentralised-moderation-pleroma}, last accessed 23 Mar 2024.}.
We capture the conversational context of each post
using GraphNLI~\cite{agarwal2022graphnli}, a state-of-the-art graph learning-based framework. This is in stark contrast to previous works~\cite{bin2022toxicity,kurita2019towards,risch2020toxic} 
that look at each post in isolation.
Because of fediverse's interoperability, conversations are scattered across instances.
We reconstruct all the conversations in Pleroma and
identify $2$ million conversations with more than two posts.
This proved to be a complex endeavour: 
we discover that each instance participating in a conversation can have a different and partial view of the complete conversation.
Indeed, interoperability poses an additional challenge: moderators may have different partial views of the conversations and therefore, may develop different perceptions about the content to be moderated.
We find that up to 14 instances participate in the same conversation and that the level of fragmentation increases with the number of instances.
We believe this challenge is inherent to interoperability.

Then we obtain toxicity scores of each of the posts in Pleroma conversations from  Perspective API. We train GraphNLI model on our Pleroma conversations for predicting these toxicity scores. We find that large  ($>$75th percentile by the number of posts) and medium instances (between 25th and 75th percentile by the number of posts) are self-sufficient: each instance can develop a model trained just on its own data and achieve a performance similar to what would be obtained by a centralised platform with access to data from all instances. 
This is a significant improvement over earlier work, which did not exploit conversational context~\cite{bin2022toxicity}.


This strategy fails with smaller instances though (\ie, $<$25th percentile). Smaller instances do not contain a sufficiently large fragment of the conversation to effectively capture the surrounding conversational context. 
We overcome this with two information-sharing strategies.
We first evaluate a \textit{data-borrowing approach}, where
small instances obtain all toots from all conversations with at least 5 toots from the largest instance they federate with. \textcolor{black}{This results in large improvements with an overall macro-F1 of $0.8258$ for small instances.}
To limit the sharing of private information, we also evaluate a \textit{model-borrowing} approach, where small instances adopt the model trained on large instances and fine-tune the model parameters. We achieve a further increase in performance ($\approx$$+3\%$ macro-F1). 

In summary, we analyse the moderation challenges resulting from decentralisation and interoperability.
In doing so, we perform the first-ever dissection of \textit{conversations} in the fediverse using data collected from Pleroma, a large fediverse micro-blogging platform.
We observe that interoperability results in heavily decentralised conversations and examine its implications on content moderation. We address the following Research Questions (RQs):

\begin{itemize}
    \item \pb{RQ1.} \emph{What do Pleroma conversations look like?}
    We reconstruct all conversations in $713$ Pleroma instances in Section~\ref{sec:online-conv} by linking posts and reposts of these posts across instances. We find over 2 million conversations that have more than two posts. 
    Interestingly, users see \textit{fragmented conversations} because different instances have different fragments of the same conversation.
    The fragmentation of conversations and the likelihood of partial federation increases with the number of instances involved in a conversation. Unfortunately, we also find that some of these larger conversations are also more likely to contain toxic content. This leads us to develop strategies to identify such content (RQ2 and RQ3).

    \item \pb{RQ2.} \emph{Can instances locally detect toxic content by exploiting conversational context?} 
    Using GraphNLI, a state-of-the-art graph-based framework~\cite{agarwal2022graphnli,agarwal2023graph}, we find 
    in Section~\ref{subsec:global-local}
    that the model trained on local content succeeds in identifying toxicity in Medium and Large instances (macro-F1 of $0.7675$ and $0.8837$,  respectively). However, we find that this method \textit{does not detect any toxic content at all for small instances}, resulting in an F1-score of 0.
    
    \item \pb{RQ3.} \emph{Can large instances help small ones in detecting toxic content?}
    We propose and evaluate  strategies  where small instances exchange information with larger ones
    in Section~\ref{subsec:federation-local}. 
    First, we explore modifications to the current federation mechanism, to federate the \textit{full conversation}. Then we propose a strategy that uses toots or posts from the largest instance that an instance federates with; we call this approach \textit{toot exchange}. Finally, we develop a more advanced approach wherein a small instance adapts the model trained by the largest instance it federates with; we call this \textit{model exchange}. While full conversation federation does not help the small instances, we find that both toot and model exchange are extremely effective, with model exchange improving the macro F1-score on small instances by up to $28.26\%$.
\end{itemize}

\section{Background \& Related Work}

\subsection{Pleroma Background}
Pleroma is an open-source decentralised and interoperable micro-blogging server platform where users create their own social communities hosted on independent servers, aka \textbf{instances}. Each Pleroma instance works similarly to Twitter, allowing users to register new accounts and share \textbf{statuses} (aka \textbf{toots}) with their followers --- equivalent to a tweet on Twitter. Users can also \textbf{repeat} others' statuses --- equivalent to retweeting on Twitter. 

In contrast to a user on Twitter, a Pleroma user on one instance can follow users from other instances as well, and Pleroma federates content from other instances as required (federation is described further below). This allows \textit{decentralised} operation and administration --- each instance can function independently of other instances. 



\pb{Interoperability.}
Decentralisation (\eg, fediverse, IPFS~\cite{10.1145/3544216.3544232}) has emerged as a way to reduce the growing concentration in the Internet~\cite{doan2022empirical}.
To further increase competition and reduce concentration~\cite{doan2022empirical,varian2019recent},
interoperability has been proposed~\cite{beck2023breaking} and will soon be mandatory for large (``gateway'') social networks in the EU thanks to the DMA~\cite{noauthor_q_nodate}.
While work is underway to implement interoperability for these large players~\cite{ietf-mimi-content-00},  
the W3C's ActivityPub subscription protocol~\cite{activitypub2018} already supports interoperability.

\pb{Fediverse.} 
The fediverse refers to the growing group of ActivityPub compatible, and therefore interconnected, platforms. 
Pleroma is thus interoperable with other services that support ActivityPub, including micro-blogging implementations such as Mastodon~\cite{zia2023flocking}. 

\pb{Federation.}
Fediverse instances are \textit{interoperable} and can \textbf{federate}: users registered on one Pleroma instance can follow users registered on another instance from Pleroma (or any other ActivityPub supporting service).
This results in the instance \textbf{subscribing} to posts performed on the remote instance, such that they can be pushed across and presented to local users.  
We refer to users registered on the same instance as \textbf{local}, and users registered on different instances as \textbf{remote}. Note that a user registered on their local instance does \emph{not} need to register with the remote instance to follow the remote user. Instead, a user just creates a single account with their local instance; when the user wants to follow a user on a  remote instance, the user's local instance performs the subscription on the user's behalf.

\subsection{Related Work}\label{sec:related-work}

\pb{Decentralised Social Networks.} 
Powered by the ActivityPub subscription protocol~\cite{activitypub2018}, multiple social networks have recently emerged and gained popularity, thanks, in part to the recent acquisition of Twitter by Elon Musk~\cite{zia2023flocking}.
Decentralisation is challenging and earlier work has found that strong re-centralisation forces exist in Mastodon~\cite{raman2019challenges}.
Recent work found that, paradoxically, while most users join the larger instances, smaller instances attract more active users~\cite{zia2023flocking}.
The general characterisation of fediverse services has focused on the content of the posts and found that user behaviour does vary across instances~\cite{la2022network,la2022information}.
In contrast, we focus on conversations rather than posts: we reconstruct Pleroma's conversations and characterise them. We find that instances frequently have a different and fragmented view of a conversation due to the partial federation of the toots that compose it.

\pb{Toxicity Detection.} 
Many studies have examined toxic activities in social networks, including for
Twitter~\cite{ribeiro2018characterizing,burnap2015cyber,waseem2016hateful}, 
Reddit~\cite{chandrasekharan2017you,mohan2017impact,almerekhi2020investigating,wu2022conversations,yin2023annobert}, and 4Chan~\cite{bernstein20114chan,papasavva2020raiders}. 
Deep learning has been commonly used to classify toxic posts~\cite{badjatiya2017deep, rui2020deephate}.
We also use a deep learning approach, however, we explicitly include conversational context.
We use GraphNLI~\cite{agarwal2022graphnli}, a recent graph-based deep learning model.
GraphNLI was originally used to infer polarities in the debating platform, Kialo. We repurpose GraphNLI for toxicity detection in a decentralised environment.

\pb{Moderation in the Fediverse.}
Several works have recently addressed the challenges of decentralised moderation.
\citet{zignani2019mastodon} explored content warnings from Mastodon,
\citet{hassan2021exploring} analysed policies in Pleroma, and
\citet{anaobi2023will} measured the heavy moderation load of instance administrators.
Similar to fediverse services, centralised platforms such as Reddit also rely on volunteer moderation~\cite{waseem2016hateful}.
Differently from the fediverse, these centralised platforms have no interoperability with other platforms.

Closest to our work is~\citet{bin2022toxicity}, which measured the presence of toxicity in Pleroma and proposed a model exchange mechanism between instances to improve local moderation capabilities.
\citet{bin2022toxicity} focuses exclusively on the 30 largest instances and individual posts.
As a result, local models perform poorly (macro-F1 score less than or equal to 0.6 for the great majority of instances) and all instances need to exchange models. 
Our conversational approach improves those results for all
713 instances, rather than just the largest 30.
The additional conversational context allows large and medium instances to be self-sufficient with similar performance to a global model. 
We also look into smaller instances.
While smaller instances typically have, by nature, a smaller moderation load~\cite{gehl2022digital}, we believe that supporting semi-automated moderation for any player -- large or small is useful. Semi-automated moderation can have potential benefits in the context of Pleroma and the fediverse: 1) federated conversations might become viral and overwhelm the administrator; 2) some administrators might be time-constrained, \eg, because they admin multiple instances~\cite{hassan2021exploring}, or because this is simply a side activity for them; 3) it might enable new use-cases or allow more people to become administrators thanks to the reduced moderation load.

\section{Problem Statement}\label{sec:problem-statement}

Interoperability and decentralisation of platforms such as Pleroma give communities the independence to exchange ideas and content without having to rely on the whims and gatekeeping of centralized platforms like Twitter.
However, the interoperable nature of fediverse platforms leads to new questions about the \textit{means by which} content can be moderated in a decentralised way, without the scale of platforms like Twitter, or when the full conversation is fragmented across instances.

\begin{figure}[!t]
\centering
\includegraphics[width=0.7\columnwidth]{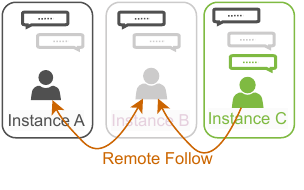}
\caption{Example scenario of a fragmented conversation in the fediverse due to partial federation: Instance C has full visibility on the conversation (3 toots) but instances A and B can only see a fragment of the conversation (2 toots)}
\label{fig:frag_conversation}
\end{figure}

\pb{Decentralised Moderation.}
The administration of fediverse instances is strictly decentralised, \ie the administrator of an instance is responsible for formulating and implementing its own moderation policy. 
However, administrators have only limited control of the content of the instance they administer, as content generated on one instance can easily spread to another one thanks to the content federation enabled by the ActivityPub protocol.
With instance administrators typically operating in a pro bono voluntary fashion, moderation and other administrative aspects of running an instance can be rather challenging.
While machine learning could in principle help to semi-automate this task, the limited amount of data on each instance constrains this possibility. 

\pb{Fragmented Conversations.}
Toots are usually part of a longer conversation involving other toots from multiple users.
This context is frequently necessary to understand the meaning and moderate content adequately.
However, the interoperable nature of the fediverse can result in a partial federation where a conversation is fragmented across instances. \textcolor{black}{Conversations can be fragmented partially due to moderation and blocking by other instances~\cite{hassan2021exploring}}. However, conversations can also be fragmented due to how federation works (even when no instance is blocked).

For example, suppose a user in instance A posts a toot. If a user in instance B follows instance A's user, they will see this post federated to instance B's. Now if this user takes part in a conversation, another user in instance C will see their conversation since the user in instance C follows instance B's user (\emph{cf.} Figure~\ref{fig:frag_conversation}). However, if the user in instance A does not follow the user in instance C, the toot of instance C's user will not be federated to instance A even though instance A's user originated the conversation.

Such \textit{partial federation} can result in \textit{fragmented conversations} where different administrators see different parts of a conversation and therefore different contexts of the content they have to moderate. For instance, in the above example, the administrator of instance A (or a machine learning model running on their behalf) might need the context from instance C in order to determine the toxicity of a post and whether a particular post is hate speech. 
Further interoperability is likely to reproduce this challenge.

\section{Dataset and Methodology}~\label{sec:methodology}

\begin{figure*}[ht!]
\centering
    \begin{subfigure}{0.33\textwidth}
        \centering
        \includegraphics[width=\columnwidth]{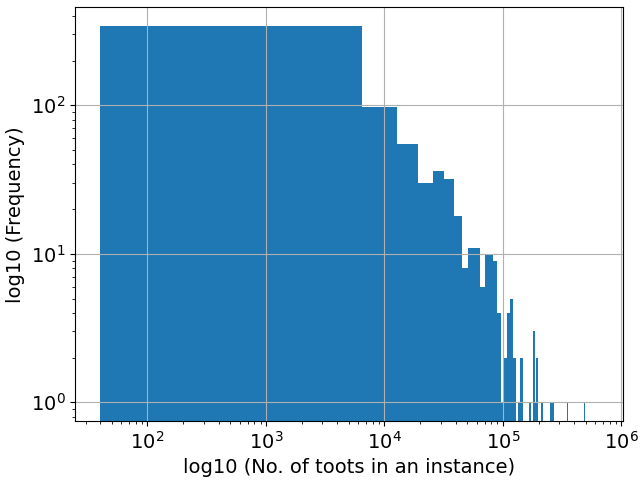}
        \caption{Histogram of the number of toots in Pleroma instances.
        }
        \label{fig:hist-num-toots}
    \end{subfigure}
    \begin{subfigure}{0.33\textwidth}
        \centering
        \includegraphics[width=\linewidth]{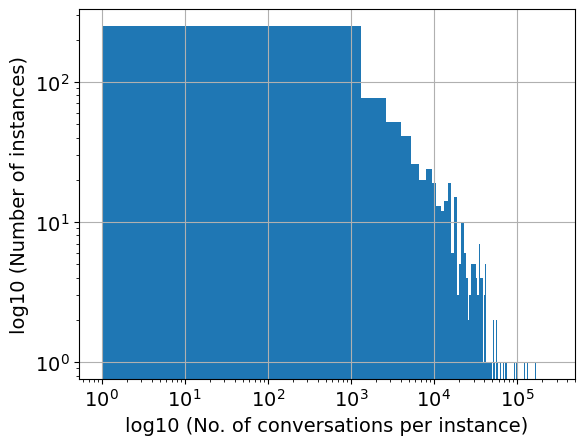}
        \caption{Histogram of number of conversations per instance.}
        \label{fig:hist-num-conv-ins}
    \end{subfigure}
    \begin{subfigure}{0.33\textwidth}
        \centering
        \includegraphics[width=\linewidth]{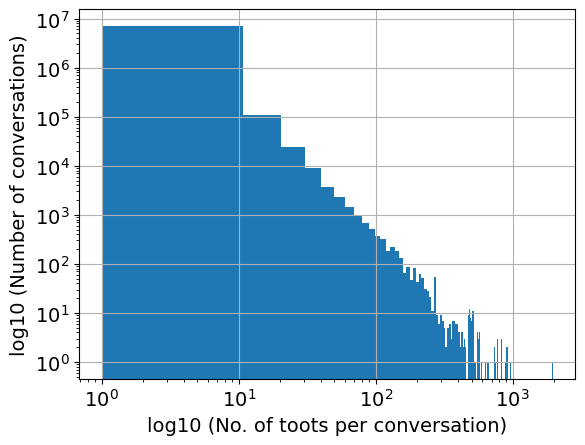}
        \caption{Histogram of the number of toots per conversation.}
        \label{fig:hist-num-toots-conv}
    \end{subfigure}
\caption{Histograms for Pleroma instances.}
\label{fig:hist-pleroma-ins}
\end{figure*}


In this section, we approach the problem of reconstructing a global, centralised view of conversations in the decentralised fediverse, and use the task of determining the toxicity of a post to investigate the impact of the fragmented conversation views on overall conversational understanding. Here, we give details of the dataset we use, and how we process it to construct conversation trees and to attach toxicity scores.

\subsection{Pleroma Dataset}
We use the Pleroma data crawled by \citet{bin2022toxicity}\footnote{\url{https://github.com/harisbinzia/mastodoner}, last accessed 1 Apr 2024.}. The dataset consists of 713 Pleroma instances with all the publicly available toots before January 22, 2021. 
In total, the dataset contains $16.5$ million toots from Pleroma, one of the largest platforms in the fediverse, by both content and user counts in the fediverse~\cite{bin2022toxicity}. The dataset represents a complete collection of all available instances at the time of data collection. Each toot in the dataset includes information about the author, text content, associated media, timestamp, number of likes, number of reblogs, and any self-tagged content warnings.

We enhance this raw dataset in two ways: First, we reconstruct the global conversations, by tracing the threads of posts and replies across instances. Second, using Google's Perspective API~\cite{perspective2023api}, we assign toxicity scores to each post in over 2 million conversations.

\subsection{Constructing Conversation Trees}
A conversation tree is a tree-structure where nodes represent the toots in conversations and a directed edge from a child to its parent node denotes that a child node replies to the parent node~\cite{agarwal2022graphnli}. A comment can have multiple replies, but it can reply to exactly one comment in Pleroma. Therefore, every node in a conversation can have only one outgoing edge (replied to) but several incoming edges (replies), thus forming a tree.

Constructing complete conversations in the fediverse is challenging as conversations can be federated across multiple instances when users from different instances participate in the same conversation. 
Partial federation can easily occur, as some toots are not federated across all the instances where a particular conversation is present.
This leads to \textit{fragmented conversations}.
We regard a ``complete conversation'' as the conversation tree that includes all toots that span as replies from the first toot in the tree.
An instance has a fragmented conversation whenever such instance can see less than $100\%$ of the complete conversation.

An obvious question is how such conversations can be reconstructed. 
Unfortunately, we find that toot IDs, parent IDs and conversation IDs are unsuitable for reconstructing conversations as these IDs are local to the instance where we observe them. 
Thus, we use the toot URL as a unique identifier for toots in the dataset as it is unique across instances. 
By mapping local toot IDs and parent IDs with toot URLs, we construct complete conversation trees which contain toots from different instances. For each node (i.e., toot), we also store the information of which instances a toot is federated to in order to study the federation of conversations. Overall, we identify $7.2$ million conversation trees in our dataset.

\subsection{Toxicity Labels}\label{sec:tox-label}
To study toxicity and moderation in Pleroma, we use toxicity scores from Google Jigsaw's Perspective API~\cite{perspective2023api} as ground truth toxicity scores, following similar studies for Pleroma~\cite{bin2022toxicity}, Mastodon~\cite{zia2023flocking}, 4chan~\cite{papasavva2020raiders} and Voat~\cite{papasavva2021qoincidence}. \textcolor{black}{We set ``doNotStore'' option to true so that the Perspective API does not store the data after scoring them for toxicity.} By extending the original dataset of~\citet{bin2022toxicity} which used toxicity scores for the top 30 instances, we obtain toxicity scores for all posts in all conversations across all instances.

The Perspective API provides a toxicity score between 0 and 1, where 1 means the highest toxicity level. \textcolor{black}{Following~\citet{bin2022toxicity,papasavva2020raiders,papasavva2021qoincidence}, we consider a toot to be toxic if its toxicity score is greater than 0.5 (moderately toxic) to capture a broad range of toxic content. For completeness, we also experimented with a stricter toxicity threshold of 0.8 and observed similar proportions of toxic toots in different Pleroma instances. To ensure the generality of our approach, we use the actual toxicity scores as the ground truth and train our models with them for the regression task of toxicity prediction.}
This ensures that our findings apply regardless of the actual threshold that an administrator might choose. More generally, we believe that Perspective labels are just one example and the approach proposed could be implemented with alternative human-based labels. Unfortunately, due to the cost-intensive nature of labelling, we do not have such an alternative.

\section{Characterising Conversations (RQ1)}\label{sec:online-conv}

Having constructed a global view of all Pleroma conversations, we next proceed to characterize them. We first perform a basic characterization to understand the distribution and size of the conversations,  then investigate the presence of toxic content, and finally develop an understanding of how individual instances or users within a conversation see the conversation, finding that conversation can be fragmented due to the nature of federation across instances.

\subsection{Basic Characterisation}
Overall, we identify $7.2$ million conversations in $713$ Pleroma instances. Figure~\ref{fig:hist-num-conv-ins} shows the histogram of the number of conversations per instance. 
On average, the number of conversations per instance is $10,380.65$, while the instance with the largest number of conversations has $267,717$ conversations. \textcolor{black}{The average number of users per instance who have at least one toot either posted or federated is $1609.60$, while the maximum number of users per instance is $9,824$.}
Next, we look at the number of toots per conversation. Figure~\ref{fig:hist-num-toots-conv} shows the histogram of the number of toots per conversation. 
We find that only $29\%$ of the conversations have $2$ or more toots. Note that even a single toot is itself a conversation, consisting of just the root of the conversation tree.
The average (median) number of toots per conversation is $2.11$ ($1$). If we only consider conversations with more than two toots, the average (median) conversation has $6.83$ ($4$) toots, whereas the maximum is $1,952$ toots.

\subsection{Toxicity in Pleroma Conversations}

\begin{figure}[!t]
\centering
\includegraphics[width=\columnwidth]{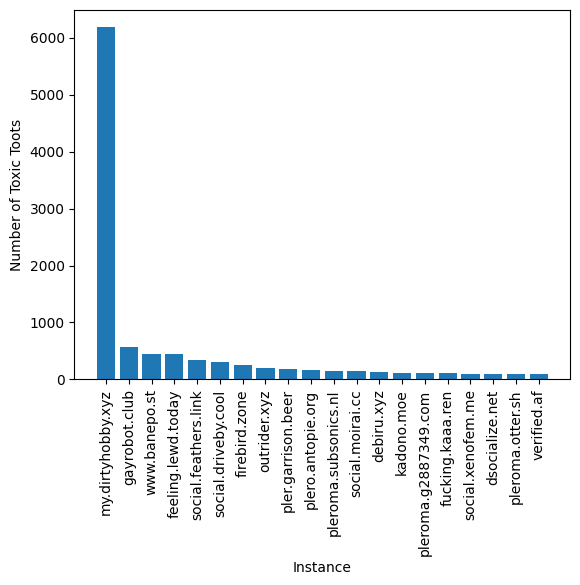}
\caption{Top 20 instances with the number of toxic toots, sorted in descending order.}
\label{fig:top-20-toxic-ins-barplot}
\end{figure}

\begin{figure*}[ht!]
\centering
    \begin{subfigure}{0.33\textwidth}
        \centering
        \includegraphics[width=\linewidth]{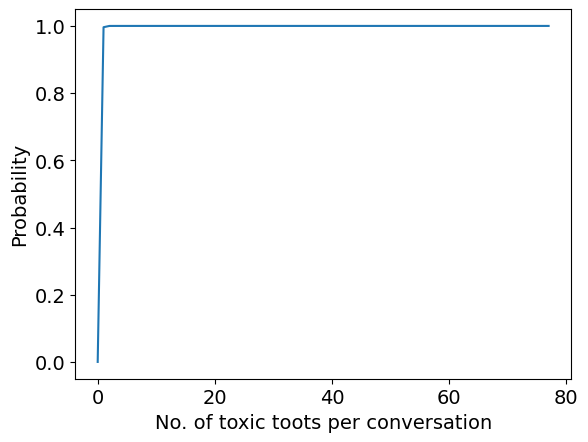}
        \caption{Number of toxic toots per conversation.
        }
        \label{fig:cdf-toxic-toots}
    \end{subfigure}
    \begin{subfigure}{0.33\textwidth}
        \centering
        \includegraphics[width=\linewidth]{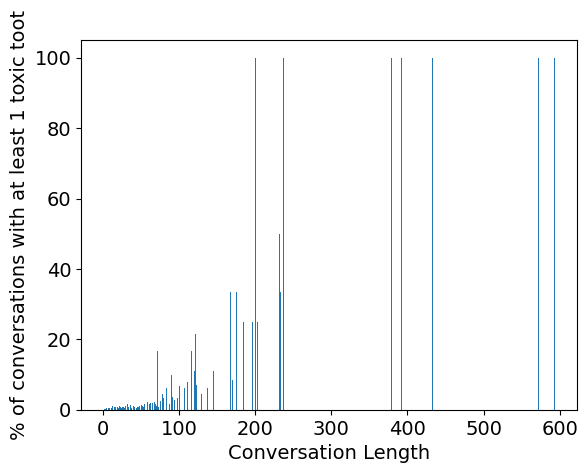}
        \caption{Percentage of conversations with at least one toxic toot versus the conversation length.}
        \label{fig:bar-toxic-toots-vs-conv-len}
    \end{subfigure}
    \begin{subfigure}{0.33\textwidth}
        \centering
        \includegraphics[width=\linewidth]{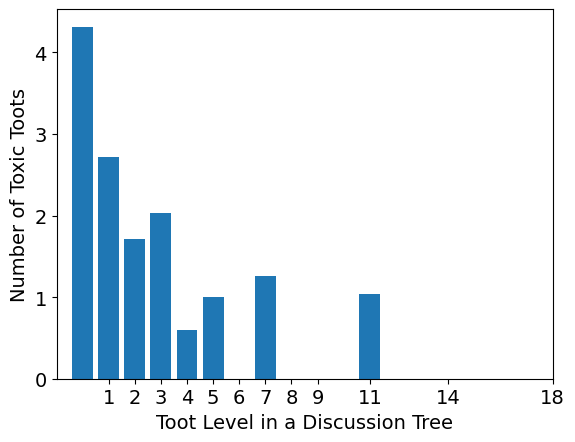}
        \caption{Number of toxic toots per toot level in a discussion tree. Y-axis is on log-scale.}
        \label{fig:bar-toot-level-toxic-toots}
    \end{subfigure}
\caption{Characterisation of Toxic Toots in Pleroma conversations.}
\label{fig:bar-plots-conv}
\end{figure*}

\pb{Presence of toxic content.}
Figure~\ref{fig:top-20-toxic-ins-barplot} shows the top 20 Pleroma instances with the number of toxic toots, sorted in descending order, indicating that a few instances are responsible for a large proportion of toxic content. \textcolor{black}{For example, \textit{my.dirtyhobby.xyz} contributes about $23.33\%$ of the toxic toots. Similarly, \textit{plero.antopie.org} and \textit{social.feathers.link} have $4.7\%$ and $3.3\%$ of the toxic content respectively.} Unfortunately, due to the nature of the federation of content, such content is not constrained solely to a few instances but instead can potentially spread more widely to other instances whose users may take part in the toxic conversations. Figure~\ref{fig:cdf-toxic-toots} shows the Cumulative Distribution Function (CDF) of the number of toxic toots per conversation. On average, the number of toxic toots per conversation is $0.004$, while the maximum number of toxic toots in a conversation can be as high as $77$, which suggests that we should be taking a conversation-centric view of toxic toots. This motivates our study in the next section, which focuses on the whole conversation context to predict toxic content.

\begin{figure*}[!ht]
\centering
    \begin{subfigure}{0.33\textwidth}
        \centering
        \includegraphics[width=\linewidth]{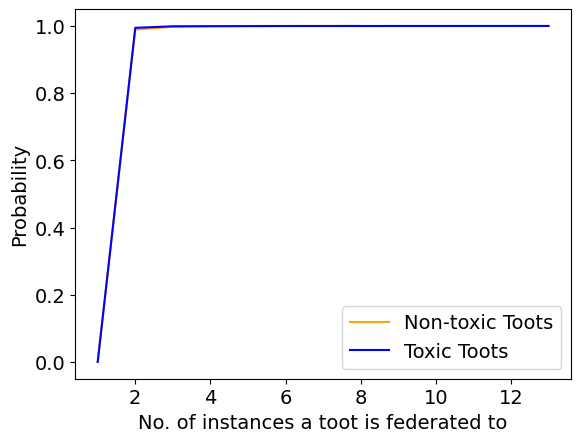}
        \caption{Number of instances toxic and non-toxic toots are federated to.}
        \label{fig:cdf-toxic-toots-fed}
    \end{subfigure}
    \begin{subfigure}{0.33\textwidth}
        \centering
        \includegraphics[width=\columnwidth]{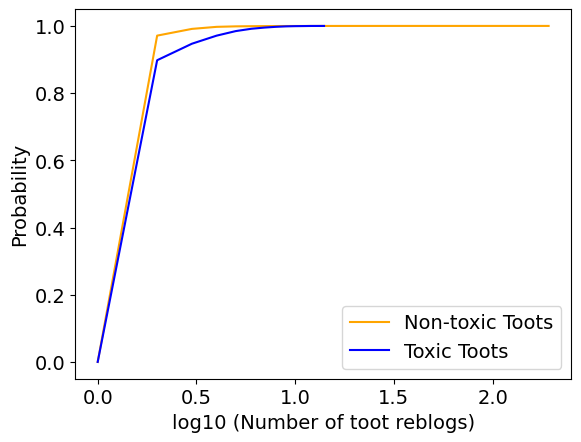}
        \caption{Number of toot reblogs for toxic and non-toxic toots.}
        \label{fig:cdf-toot-reblogs}
    \end{subfigure}
    \begin{subfigure}{0.33\textwidth}
        \centering
        \includegraphics[width=\columnwidth]{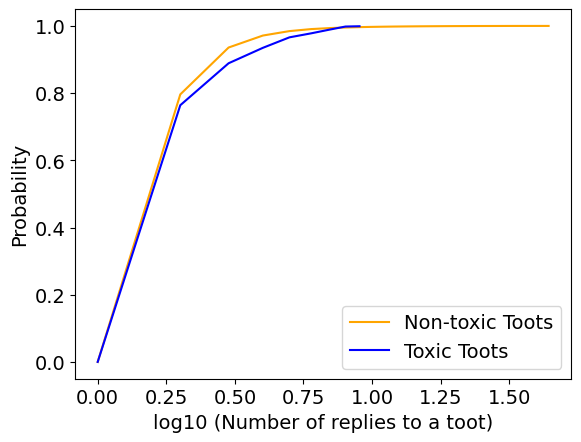}
        \caption{Number of direct replies to toxic and non-toxic toots.}
        \label{fig:cdf-toot-replies}
    \end{subfigure}
\caption{Cumulative Distribution Functions (CDFs) for toxic toots in Pleroma instances.}
\label{fig:cdfs-toxicity}
\end{figure*}

\pb{Popularity of toxic content.}
We next examine the relative popularity of toxic versus non-toxic content from different angles. First, Figure~\ref{fig:cdf-toxic-toots-fed} shows that toxic toots have a higher likelihood of ``going viral'' across multiple instances:
Whereas most toots (both toxic and non-toxic) are federated to a handful of instances, the maximum number of instances a toxic toot is federated to is $13$, while the maximum is $8$ for non-toxic toots.

Next, we compare the engagement metrics of toxic versus non-toxic toots. Figure~\ref{fig:cdf-toot-reblogs} shows the CDF of toot reblogs (reshares) for toxic and non-toxic toots. On average, the number of reblogs that toxic toots receive is $0.556$, whereas non-toxic toots receive $0.271$ reblogs. This suggests that toxic toots are reblogged (re-shared) $2$x more than the non-toxic toots in Pleroma conversations. 
Figure~\ref{fig:cdf-toot-replies} also shows the CDF of the number of direct replies to toxic versus non-toxic toots in Pleroma conversations. The average number of replies to toxic toots is $1.482$, while the average number of replies to non-toxic toots is $1.354$. Thus, toxic toots receive a higher number of replies than non-toxic toots. 

Collectively the above results clearly show that toxic toots receive higher levels of engagement (reblogs and replies) in Pleroma. Toxic content may spread faster (i.e., have more views overall) because they receive a higher number of reblogs ($2$x) than the non-toxic toots.

\pb{Where can we find toxic content?}
Finally, we examine which kind of conversations are more likely to contain toxic content, and within such conversations, where are the toxic toots located.

Figure~\ref{fig:bar-toxic-toots-vs-conv-len} shows a bar plot of the percentage of conversations with at least $1$ toxic toot versus the conversation length. As the conversation length increases, the likelihood of having toxic toots also increases. Thus, the longer a conversation, the more likely it is to have some toxic statements. 

We look deeper and ask where, within those toxic conversations, are toxic toots to be found. Figure~\ref{fig:bar-toot-level-toxic-toots} shows the bar plot of the number of toxic toots per toot level in a conversation tree. We define the ``level'' of a toot as follows: The root node in a conversation tree is at level $0$; its immediate children are at level $1$. We proceed in this fashion with increasing levels until we reach the leaf nodes. From the figure, we see that it is more likely to have toxic toots at the beginning of the conversation near the root node and other ancestor nodes. This suggests that further down in a conversation (closer to the leaves), the immediate context or content of a toot may be insufficient to detect toxicity and it may require context from further up the conversation tree. This motivates our adoption in Section~\ref{sec:tox-detection} of a graph-walk based approach~\cite{agarwal2022graphnli} that can gather context from farther up the conversation tree to classify toxicity. 

\subsection{Federation and Fragmented Conversations}\label{sec:partial-federation}

\begin{figure*}[ht!]
\centering
    \begin{subfigure}{0.33\textwidth}
        \centering
        \includegraphics[width=\columnwidth]{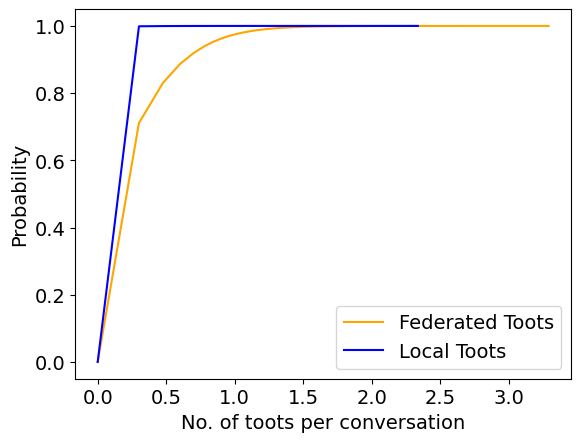}
        \caption{Number of local and federated toots in Pleroma instances (log-scale on X-axis).
        }
        \label{fig:cdf-local-fed-toots}
    \end{subfigure}
    \begin{subfigure}{0.33\textwidth}
        \centering
        \includegraphics[width=\columnwidth]{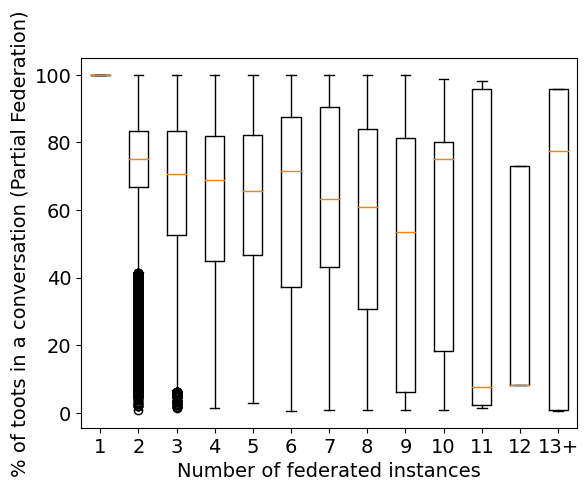}
        \caption{Box plot with the variation in the percentage of federated toots per conversation for the instances involved in Pleroma conversations.
        }
        \label{fig:partial-fed-box-plot}
    \end{subfigure}
    \begin{subfigure}{0.33\textwidth}
        \centering
        \includegraphics[width=\columnwidth]{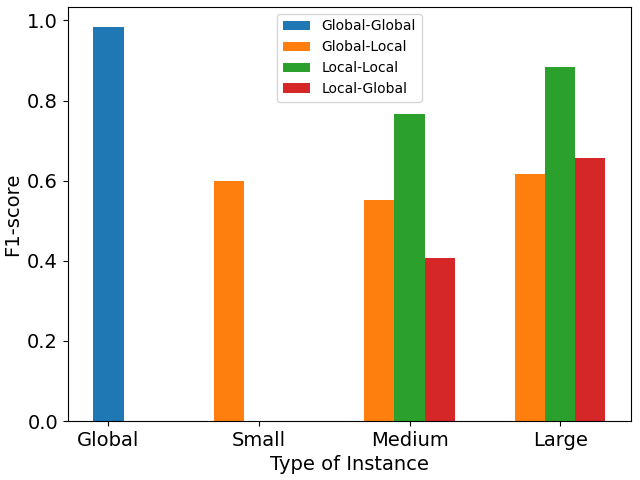}
        \caption{Bar plot with F1-scores for different types of Pleroma instances and GraphNLI models.}
        \label{fig:f1-score-barplot}
    \end{subfigure}
\caption{Conversations in Pleroma and F1-scores for toxicity detection.}
\label{fig:toxicity-figs}
\end{figure*}

Content federation is at the very essence of the fediverse and allows users to interact with each other regardless of the instance they are signed up with.
On average, a Pleroma instance federates with $17.73$ instances, while the maximum number of instances an instance federates with is $118$.
Figure~\ref{fig:cdf-local-fed-toots} shows the CDF of the number of local and federated toots per conversation in Pleroma instances. Note, local toots within an instance are the toots which originated in that instance, while federated toots are the ones which are federated from other instances. 
The average number of local toots per conversation is $0.018$, while the average number of federated toots per conversation is $2.091$. Therefore, on average $99\%$ of the toots in a conversation are federated and the majority of the toots in a Pleroma instance are federated as well. 
Next, we look at how many instances a conversation is federated to, irrespective of where the conversation has originated.
On average, the number of instances to which a conversation is federated is $1.009$, 
whereas the maximum number of instances in a conversation goes up to $14$.

Since conversations are composed of toots from different instances, partial federation is likely to result in fragmented conversations, as discussed in Section~\ref{sec:problem-statement}. Figure~\ref{fig:partial-fed-box-plot} shows how partial federation impacts the visibility of an instance on conversations depending on the number of instances involved in such conversation.
The boxplot represents the variation in the percentage of federated toots per conversation for the instances involved in conversations in Pleroma. 
The x-axis shows the number of instances involved in a conversation, and the y-axis is the percentage of toots out of the complete conversation ($100\%$ toots) that are federated to a given instance. 
A conversation is fragmented whenever partial federation takes place, i.e., when the number of toots in a conversation is below $100\%$. \textcolor{black}{It can be partially due to moderation and blocking of other instances. However, conversations can also be fragmented due to how federation works.} It can occur when users from different instances take part in a fediverse conversation, as discussed in Section~\ref{sec:problem-statement}.
We observe from Figure~\ref{fig:partial-fed-box-plot} that the fragmentation of conversations and its likelihood increases with the increase in the number of instances involved in a conversation. 

Such partial views of a conversation are likely to hamper the ability of an individual instance to discover toxic content, especially since, as detailed above, context from farther up in a conversation tree is important.

\section{Decentralised Toxicity Detection}~\label{sec:tox-detection}
%
In Section~\ref{sec:online-conv}, we have shown that toxicity is common in  Pleroma and that it spreads faster and has more engagement (in terms of reblogs and replies) than non-toxic content. It is a challenging task to not only monitor a given instance's toots but also the federated toots coming from other instances. In this section, we explore the feasibility of individual instances being able to protect themselves from toxic content using automated machine learning-based approaches. Importantly, Section~\ref{sec:online-conv} also showed that context from farther up in a conversation can potentially be relevant when labelling a post as toxic or not.

\subsection{Does Global Conversational Context Help in Detecting Toxic Content?}
The characterisation in Section~\ref{sec:online-conv} suggests that, to infer the toxicity of a toot, understanding its conversational context is beneficial.  
To test whether this holds true in practice, we adapt GraphNLI~\cite{agarwal2022graphnli}, which uses graph walks to capture the conversational context.

GraphNLI~\cite{agarwal2022graphnli} is a graph-based deep learning model which uses graph walks to capture the conversational context. In contrast to its original purpose to measure polarity, we adapt the model for regression tasks to infer toxicity scores. Using our constructed conversation trees, we first employ biased root-seeking random walks for each of the nodes in a tree to capture the relevant conversational context. A root-seeking random walk is a random walk which is biased towards the root as it selects the parent and other ancestor nodes with a higher probability than the sibling and child nodes in a conversation tree~\cite{agarwal2022graphnli}. As shown earlier, context from farther up in the conversation tree may be important for labelling toxic posts. We input the toots along with the conversational context into the GraphNLI model for predicting toxicity scores. The model is trained for $3$ epochs with an Adam optimiser.

We compare GraphNLI with the popular transformer-based BERT~\cite{devlin2018bert} that has no conversational context, i.e., the BERT model exclusively relies on the specific toot text for which it is inferring the toxicity. 
To assess which model performs better, we train both models with the complete Pleroma dataset.
For generality,  we model toxicity detection as a regression task that predicts the toxicity score that would be assigned by the perspective API for that post. Then, we convert the predicted scores into a classification (hate speech or not), based on whether the predicted toxicity score is greater than or less than the threshold of 0.5, as previously used in various studies and described in Section~\ref{sec:tox-label}.

\textcolor{black}{In all the models both here and subsequently in this paper, we split our data into 80:20 ratios for training and testing. We use a single $24$ GB Nvidia Titan RTX GPU for training and each epoch took about an hour to train GraphNLI on the entire training set. We use macro-F1 scores (due to dataset imbalance) along with accuracy to measure the performance of the classification models, and the mean-squared error (MSE) to assess the regression task. Note that the lower the MSE loss, the better the performance, whereas in the case of accuracy and macro-F1, the higher the better.}


\begin{table}[htb]
    \centering
    \begin{tabular}{l|ccc}
    \hline
        \textbf{Model} & \textbf{MSE} & \textbf{Accuracy} & \textbf{macro-F1} \\
    \hline
        BERT & 0.0686 & 0.8487 & 0.4591 \\
        GraphNLI & \textbf{0.0056} & \textbf{0.9744} & \textbf{0.9846} \\
    \hline
    \end{tabular}
    \caption{Performance of machine learning models for toxicity detection trained on the complete Pleroma dataset.}
    \label{tab:baselines}
\end{table}

Table~\ref{tab:baselines} shows the performance of both models. Our model with conversational context, GraphNLI, clearly outperforms the BERT baseline, with an overall MSE of 0.0056 and 0.9846 macro-F1 score.



\subsection{Global versus Local Moderation (RQ2)}
~\label{subsec:global-local}

\begin{table*}[htb]
\centering
\scalebox{0.95}{
\begin{tabular}{l|ccc|ccc|ccc}
\hline
\textbf{GraphNLI} & \multicolumn{3}{c}{Small} & \multicolumn{3}{c}{Medium} & \multicolumn{3}{c}{Large}    \\
\hline
\textbf{Training-Inference} & \textbf{MSE} & \textbf{Accuracy} & \textbf{macro-F1} & \textbf{MSE} & \textbf{Accuracy} & \textbf{macro-F1} & \textbf{MSE} & \textbf{Accuracy} & \textbf{macro-F1}     \\                  
\hline
Global-Global & 0.0056 & 0.9744 & 0.9846 & 0.0056 & 0.9744 & 0.9846 & 0.0056 & 0.9744 & 0.9846   \\
\hline
Global-Local & 0.0506 & 0.8726 & \textbf{0.60} & 0.0461 & 0.9093 & 0.5513 & 0.0502 & 0.8637 & 0.6162   \\
Local-Global & 0.2781 & 0.3230 & 0.0 & 0.1758 & 0.4358 & 0.4081 & 0.0781 & 0.6565 & 0.6557   \\
Local-Local & \textbf{0.0091} & \textbf{0.9743} & 0.0 & \textbf{0.0076} & \textbf{0.9729} & \textbf{0.7675} & \textbf{0.0038} & \textbf{0.9731} & \textbf{0.8837}  \\
\hline
\end{tabular}
}
\caption{Toxicity detection performance of GraphNLI for Global v/s Local models at Pleroma instances of different sizes.}
\label{tab:results}
\end{table*}

We term the above model, which uses a centralized global view of the entire data, as our \textbf{global} model. While the performance here is very robust, it is also infeasible in practice as individual instances will not have such a global view of data. 

The decentralised moderation that characterises the fediverse implies that any form of administrative task or automation can only rely on local data. The partial federation and fragmented conversations discovered in the previous section may therefore present a barrier to effective automated models. To test the limitations posed by decentralization, we develop models that can work \textbf{locally} in each Pleroma instance using the data available just on that instance, and compare the results with our global model. 

As we expect model performance to vary depending on the amount of training data available, we divide the training data of the Pleroma instances depending on the instance size (i.e., according to the number of toots).
We classify instances as \textit{Small}, \textit{Medium}, and \textit{Large}, depending on whether they are in the $25th$ quantile, between $25th$ and $75th$ quantile, or greater than $75th$ quantile, respectively.
Figure~\ref{fig:hist-num-toots} shows the histogram of the number of toots in Pleroma instances. In the distribution of toots, the $25th$ quantile is $1,290$ toots, the $50th$ quantile is $7,020$ toots, and the $75th$ quantile is $26,960$ toots; thus large instances are an order of magnitude larger than the small instances.

Table~\ref{tab:results} shows the toxicity detection performance of GraphNLI for \textit{Global} and \textit{Local} models for Pleroma instances of different sizes. \textcolor{black}{We report average results for each instance type based on the size and find an overall standard deviation of $0.89$.} Global-Global means that the model is trained globally in a centralised manner and inferred globally as well; thus creating a situation akin to today's centralised platforms such as Twitter, which have all the data in one location, allowing them to train large and accurate models. In the decentralised version, however, we can have other combinations:  Global-Local refers to global training and local inference (i.e., the random walks of GraphNLI have access to conversation context from all instances during training, but during inference, the walks can only use the local instance, and thus have access to only the local context).
This is more realistic than global training and global inference, as data exchange across instances only needs to happen once during training; rather than during inference, which happens for every toot. Similarly, the local-global model involves training using the local instance data and using the global data (i.e., the conversation across all instances involved) for inference. The local-global model needs global data for every inference made (i.e., for every new post/toot), and so is less feasible than the global-local model which uses global data only during training. However, this mechanism is shown for completeness. Finally, the local-local model is the most realistic, as it uses only the local instance for both training and inference.


Figure~\ref{fig:f1-score-barplot} shows the F1-scores obtained for all combinations. The global training-global inference model is a single model representing a centralised equivalent of the decentralised fediverse, only shown here as a benchmark. When there is local training or local inference, there is per-instance variation in performance. We show this as separate bar plots for instances of different sizes -- small ($<$25th percentile in number of toots), medium ($>$25th percentile and $<$75th percentile) and large ($>$75th percentile). For the small instances, local training does not work at all; and we are unable to detect \textit{any} hate speech (i.e., a toot with toxicity $>0.5$).
On the other hand, for medium and large instances, local training and local inference appear to perform better than variants involving either global training or global inference; suggesting that local context, when sufficiently rich, is preferable to global context from farther away on a different instance.
%
Global training Local inference seems to be the only choice with macro-F1 of $0.60$ for moderating Small instances. But the performance of the model on Small instances is not good enough and most importantly, Global training violates the principles of the decentralised web, which is based on each instance being as independent as possible. Hence, we further investigate how we can improve the local moderation of Small instances.

\subsection{Federation Strategies to Improve Local Moderation (RQ3)}
~\label{subsec:federation-local}

\begin{table*}[]
    \centering
    \scalebox{0.95}{
    \begin{tabular}{l|ccc|ccc|ccc}
    \hline
    \textbf{Local-Local} & \multicolumn{3}{c}{Small} & \multicolumn{3}{c}{Medium} & \multicolumn{3}{c}{Large}    \\
    \hline
    \textbf{Federation Strategy} & \textbf{MSE} & \textbf{Accuracy} & \textbf{macro-F1} & \textbf{MSE} & \textbf{Accuracy} & \textbf{macro-F1} & \textbf{MSE} & \textbf{Accuracy} & \textbf{macro-F1}     \\
    \hline
        Full Conversations & 0.0230 & 0.9481 & 0.0 & 0.0073 & 0.9654 & 0.7696 & 0.0041 & 0.9711 & 0.8826   \\
        Toot Federation & \textbf{0.0063} & 0.9678 & 0.8258 & 0.0067 & 0.9747 & 0.8077 & 0.0040 & 0.9813 & 0.8841    \\
        Model Sharing & 0.0075 & \textbf{0.9786} & \textbf{0.8826} & \textbf{0.0061} & \textbf{0.9773} & \textbf{0.8169} & \textbf{0.0038} & \textbf{0.9839} & \textbf{0.8872}   \\
    \hline
    \end{tabular}
    }
    \caption{Performance of GraphNLI model with different federation strategies for local content moderation of instances of different sizes.}
    \label{tab:toot-fed-results}
\end{table*}


We demonstrated in the previous section that local moderation is effective for large and medium-sized instances.
Unfortunately, this strategy fails for Small instances due to the limited amount of data (and in particular toxic data). 
While the focus is usually on the larger instances, ensuring that smaller ones are viable can be critical to reducing the barrier of entry for new players and promoting the long-term sustainability of the ecosystem.

We now devise strategies to solve this problem.
We experiment with three federation strategies to improve the local moderation: 
\one \textbf{Full Conversations:} we federates toots from other instances to complete the fragmented conversations (due to partial federation), thereby increasing the training data and conversational context for our GraphNLI model; 
\two \textbf{Toot Federation:} we select the largest instance that federates with the target instance and then federates all of the toots in conversations with a length of more than $5$.
\three \textbf{Model Sharing:} we share the trained toxicity detection GraphNLI model from the largest federating instance with the target instance. The target instance then fine-tunes on top of the already trained model using its local conversations.

Table~\ref{tab:toot-fed-results} shows the performance of the GraphNLI model with Local training and Local inference with different federation strategies for local content moderation.
We observe that even with full conversations, macro-F1 is still $0$ for Small instances since no toxic toots are predicted correctly because of small datasets available locally.
Toot federation with the largest federating instance works well with an overall accuracy of $0.9678$ and $0.8258$ macro-F1 score for Small instances.
The model sharing strategy performs the best with an accuracy of $0.9786$ and $0.8826$ macro-F1 score. This represents an improvement of $28.26\%$ in macro-F1 score over our previous best-performing variant (Global-Local) for Small instances.
Model sharing improves the local moderation of Small instances in a privacy-preserving way since no actual toots content needs to be shared. However, it requires a communication channel with high bandwidth to share the weights of deep learning models between Pleroma instances. 


In contrast, the toot federation strategy is feasible through existing channels between instances, since these instances continuously federate content among themselves. For instance, an administrator could simply choose to follow a specific account from a remote instance it is already federating with to ensure additional toot federation.
We also experiment with these federation strategies on Medium and Large instances as shown in Table~\ref{tab:toot-fed-results}. For Medium instances, the macro-F1 score goes up to $0.8169$ for the model-sharing strategy, which is an improvement of about $5\%$. We find that there is only a slight improvement for Large instances because they already gain good performance on toxicity detection through local training only due to the large training dataset. Therefore, Medium and Large instances may not have strong incentives for federated moderation and may choose to deal with moderating toxic content locally. However, toot federation or model sharing is very beneficial for small instances. \textcolor{black}{Since the model-sharing strategy performs the best for small instances, and only requires sharing the model weights (no text or toots), model-sharing can be the preferred strategy with better privacy-preserving properties.
}

\subsection{Limitations and Ethical Considerations}

The data includes user information and public posts. 
In collecting it, we took an ``algorithmic thinking in the public interest'' approach~\cite{luscombe2022algorithmic}.
We believe that our study has large societal benefits that outweigh potential risks: content moderation remains a challenge. We believe that research into novel moderation scenarios and approaches has the potential for large societal benefits that expand from Pleroma and the fediverse to other platforms, particularly if they are interoperable. The benefits include: \one scrutiny into current moderation practices; \two identification of problems; \three potential solutions to help address them; and \four timeliness, as this is a young ecosystem with very different moderation and interaction mechanisms that are currently understudied but have a large potential to become mainstream. 

We exclusively use publicly accessible information, adhering to well-established ethical protocols for collecting social data. \textcolor{black}{We obtained Ethics approval from both the home institution of one of the authors, and independently from REPHRAIN, UK’s National Research Center for Online Harms}. The authors also declare that they do not have any competing interests.

We also strive to minimise potential harm~\cite{fiesler2016exploring} and anticipate minimal risk: \one we were careful to not overload the servers; \two the data collected is publicly available; \three the data is collected through public APIs; \four we anonymise the data upon collection; \five we do not plan to make data publicly available; \six we ensure that no data is stored on the Perspective API servers (we choose the “doNotStore” option\footnote{\url{https://support.perspectiveapi.com/s/about-the-api-faqs?language=en_US)}, last accessed 14 Sep 2023.}).

\section{Conclusions}~\label{sec:concl}
This paper proposed and evaluated novel moderation solutions for decentralised, interoperable social networks. We believe this is particularly needed because interoperability is likely to increase thanks to regulation (DMA~\cite{noauthor_q_nodate}) and the growing support for ActivityPub.
We leverage the conversational context of each post with GraphNLI, a graph-based deep learning framework that makes random walks on the posts of the conversation.
We evaluated our approaches using a comprehensive dataset of Pleroma (consisting of $713$ instances and $16.5$ million posts), a large decentralised micro-blogging service of the fediverse.
We first reconstructed the $2$ million conversations with more than two posts that exist in our data.
We found this to be more complex than expected as instances participating in the same conversation might only have a partial view of the entire conversation.
Our conversation-based approach, succeeded in detecting toxicity in large instances ($0.8837$ macro-F1) but failed for small ones.
We solve this problem via federation: we evaluate different strategies for sharing information across instances (toots or models) and achieve the highest macro-F1 of $0.8826$ for model sharing strategy in small instances.


\section*{Acknowledgements}
This work was supported by AP4L (EP/W032473/1), EU Horizon Framework grant agreement 101093006 (TaRDIS) and REPHRAIN’s ``DSNMod'' and ``Fediobservatory'' (EP/V011189/1).


\bibliography{aaai24}

\begin{thebibliography}{39}
\providecommand{\natexlab}[1]{#1}

\bibitem[{Agarwal et~al.(2022)Agarwal, Joglekar, Young, and
  Sastry}]{agarwal2022graphnli}
Agarwal, V.; Joglekar, S.; Young, A.~P.; and Sastry, N. 2022.
\newblock GraphNLI: A Graph-based Natural Language Inference Model for Polarity
  Prediction in Online Debates.
\newblock In \emph{Proceedings of the ACM Web Conference 2022}, 2729--2737.

\bibitem[{Agarwal et~al.(2023)Agarwal, Young, Joglekar, and
  Sastry}]{agarwal2023graph}
Agarwal, V.; Young, A.~P.; Joglekar, S.; and Sastry, N. 2023.
\newblock A graph-based context-aware model to understand online conversations.
\newblock \emph{ACM Transactions on the Web}, 18(1): 1--27.

\bibitem[{Almerekhi, Jansen, and Kwak(2020)}]{almerekhi2020investigating}
Almerekhi, H.; Jansen, B.~J.; and Kwak, H. 2020.
\newblock Investigating toxicity across multiple Reddit communities, users, and
  moderators.
\newblock In \emph{Companion proceedings of the web conference 2020}, 294--298.

\bibitem[{Anaobi et~al.(2023)Anaobi, Raman, Castro, Zia, Ibosiola, and
  Tyson}]{anaobi2023will}
Anaobi, I.~H.; Raman, A.; Castro, I.; Zia, H.~B.; Ibosiola, D.; and Tyson, G.
  2023.
\newblock Will Admins Cope? Decentralized Moderation in the Fediverse.
\newblock In \emph{Proceedings of the ACM Web Conference 2023}, 3109--3120.

\bibitem[{Badjatiya et~al.(2017)Badjatiya, Gupta, Gupta, and
  Varma}]{badjatiya2017deep}
Badjatiya, P.; Gupta, S.; Gupta, M.; and Varma, V. 2017.
\newblock Deep Learning for Hate Speech Detection in Tweets.
\newblock In \emph{Proceedings of the 26th International Conference on World
  Wide Web Companion}, WWW '17 Companion, 759–760. Republic and Canton of
  Geneva, CHE: International World Wide Web Conferences Steering Committee.
\newblock ISBN 9781450349147.

\bibitem[{Beck and Moore(2023)}]{beck2023breaking}
Beck, M.~D.; and Moore, T.~R. 2023.
\newblock Breaking Up a Digital Monopoly.
\newblock \emph{Communications of the ACM}, 66(6): 38--41.

\bibitem[{Bernstein et~al.(2011)Bernstein, Monroy-Hern{\'a}ndez, Harry,
  Andr{\'e}, Panovich, and Vargas}]{bernstein20114chan}
Bernstein, M.; Monroy-Hern{\'a}ndez, A.; Harry, D.; Andr{\'e}, P.; Panovich,
  K.; and Vargas, G. 2011.
\newblock {4chan and/b: An Analysis of Anonymity and Ephemerality in a Large
  Online Community}.
\newblock In \emph{Proceedings of the International AAAI Conference on Web and
  Social Media}, volume~5.

\bibitem[{Bin~Zia et~al.(2022)Bin~Zia, Raman, Castro, Hassan~Anaobi,
  De~Cristofaro, Sastry, and Tyson}]{bin2022toxicity}
Bin~Zia, H.; Raman, A.; Castro, I.; Hassan~Anaobi, I.; De~Cristofaro, E.;
  Sastry, N.; and Tyson, G. 2022.
\newblock Toxicity in the decentralized web and the potential for model
  sharing.
\newblock \emph{Proceedings of the ACM on Measurement and Analysis of Computing
  Systems}, 6(2): 1--25.

\bibitem[{Birman(2018)}]{birman2018moderation}
Birman, I. 2018.
\newblock Moderation in different communities on Reddit--A qualitative analysis
  study.

\bibitem[{Burnap and Williams(2015)}]{burnap2015cyber}
Burnap, P.; and Williams, M.~L. 2015.
\newblock Cyber hate speech on twitter: An application of machine
  classification and statistical modeling for policy and decision making.
\newblock \emph{Policy \& internet}, 7(2): 223--242.

\bibitem[{Cao, Lee, and Hoang(2020)}]{rui2020deephate}
Cao, R.; Lee, R. K.-W.; and Hoang, T.-A. 2020.
\newblock DeepHate: Hate Speech Detection via Multi-Faceted Text
  Representations.
\newblock In \emph{12th ACM Conference on Web Science}, WebSci '20, 11–20.
  New York, NY, USA: Association for Computing Machinery.
\newblock ISBN 9781450379892.

\bibitem[{Chandrasekharan et~al.(2017)Chandrasekharan, Pavalanathan,
  Srinivasan, Glynn, Eisenstein, and Gilbert}]{chandrasekharan2017you}
Chandrasekharan, E.; Pavalanathan, U.; Srinivasan, A.; Glynn, A.; Eisenstein,
  J.; and Gilbert, E. 2017.
\newblock You can't stay here: The efficacy of reddit's 2015 ban examined
  through hate speech.
\newblock \emph{Proceedings of the ACM on Human-Computer Interaction}, 1(CSCW):
  1--22.

\bibitem[{Devlin et~al.(2018)Devlin, Chang, Lee, and
  Toutanova}]{devlin2018bert}
Devlin, J.; Chang, M.-W.; Lee, K.; and Toutanova, K. 2018.
\newblock Bert: Pre-training of deep bidirectional transformers for language
  understanding.
\newblock \emph{arXiv preprint arXiv:1810.04805}.

\bibitem[{Doan et~al.(2022)Doan, van Rijswijk-Deij, Hohlfeld, and
  Bajpai}]{doan2022empirical}
Doan, T.~V.; van Rijswijk-Deij, R.; Hohlfeld, O.; and Bajpai, V. 2022.
\newblock An empirical view on consolidation of the web.
\newblock \emph{ACM Transactions on Internet Technology (TOIT)}, 22(3): 1--30.

\bibitem[{EU(2023)}]{noauthor_q_nodate}
EU. 2023.
\newblock Q\&{A}: {DMA}: {Ensuring} fair and open digital markets.

\bibitem[{Fiesler et~al.(2016)Fiesler, Wisniewski, Pater, and
  Andalibi}]{fiesler2016exploring}
Fiesler, C.; Wisniewski, P.; Pater, J.; and Andalibi, N. 2016.
\newblock Exploring ethics and obligations for studying digital communities.
\newblock In \emph{Proceedings of the 2016 ACM International Conference on
  Supporting Group Work}, 457--460.

\bibitem[{Gehl and Zulli(2022)}]{gehl2022digital}
Gehl, R.~W.; and Zulli, D. 2022.
\newblock The digital covenant: non-centralized platform governance on the
  mastodon social network.
\newblock \emph{Information, Communication \& Society}, 1--17.

\bibitem[{Hassan et~al.(2021)Hassan, Raman, Castro, Zia, De~Cristofaro, Sastry,
  and Tyson}]{hassan2021exploring}
Hassan, A.~I.; Raman, A.; Castro, I.; Zia, H.~B.; De~Cristofaro, E.; Sastry,
  N.; and Tyson, G. 2021.
\newblock Exploring content moderation in the decentralised web: The pleroma
  case.
\newblock In \emph{Proceedings of the 17th International Conference on emerging
  Networking EXperiments and Technologies}, 328--335.

\bibitem[{He et~al.(2023)He, Zia, Castro, Raman, Sastry, and
  Tyson}]{zia2023flocking}
He, J.; Zia, H.~B.; Castro, I.; Raman, A.; Sastry, N.; and Tyson, G. 2023.
\newblock Flocking to mastodon: Tracking the great twitter migration.
\newblock In \emph{Proceedings of the 2023 ACM on Internet Measurement
  Conference}, 111--123.

\bibitem[{Iqbal et~al.(2022)Iqbal, Arshad, Tyson, and
  Castro}]{iqbal2022exploring}
Iqbal, W.; Arshad, M.~H.; Tyson, G.; and Castro, I. 2022.
\newblock Exploring crowdsourced content moderation through lens of reddit
  during covid-19.
\newblock In \emph{Proceedings of the 17th Asian Internet Engineering
  Conference}, 26--35.

\bibitem[{Jigsaw(2023)}]{perspective2023api}
Jigsaw, G. 2023.
\newblock Perspective API.
\newblock \url{https://www.perspectiveapi.com/}.
\newblock Accessed: 2024-04-08.

\bibitem[{Kurita, Belova, and Anastasopoulos(2019)}]{kurita2019towards}
Kurita, K.; Belova, A.; and Anastasopoulos, A. 2019.
\newblock Towards robust toxic content classification.
\newblock \emph{arXiv preprint arXiv:1912.06872}.

\bibitem[{La~Cava, Greco, and
  Tagarelli(2022{\natexlab{a}})}]{la2022information}
La~Cava, L.; Greco, S.; and Tagarelli, A. 2022{\natexlab{a}}.
\newblock Information consumption and boundary spanning in decentralized online
  social networks: the case of mastodon users.
\newblock \emph{Online Social Networks and Media}, 30: 100220.

\bibitem[{La~Cava, Greco, and Tagarelli(2022{\natexlab{b}})}]{la2022network}
La~Cava, L.; Greco, S.; and Tagarelli, A. 2022{\natexlab{b}}.
\newblock Network analysis of the information consumption-production dichotomy
  in mastodon user behaviors.
\newblock In \emph{Proceedings of the International AAAI Conference on Web and
  Social Media}, volume~16, 1378--1382.

\bibitem[{Luscombe, Dick, and Walby(2022)}]{luscombe2022algorithmic}
Luscombe, A.; Dick, K.; and Walby, K. 2022.
\newblock Algorithmic thinking in the public interest: navigating technical,
  legal, and ethical hurdles to web scraping in the social sciences.
\newblock \emph{Quality \& Quantity}, 56(3): 1023--1044.

\bibitem[{Mahy(2023)}]{ietf-mimi-content-00}
Mahy, R. 2023.
\newblock {More Instant Messaging Interoperability (MIMI) message content}.
\newblock Internet-Draft draft-ietf-mimi-content-00, Internet Engineering Task
  Force.
\newblock Work in Progress.

\bibitem[{Mohan et~al.(2017)Mohan, Guha, Harris, Popowich, Schuster, and
  Priebe}]{mohan2017impact}
Mohan, S.; Guha, A.; Harris, M.; Popowich, F.; Schuster, A.; and Priebe, C.
  2017.
\newblock The impact of toxic language on the health of reddit communities.
\newblock In \emph{Canadian Conference on Artificial Intelligence}, 51--56.
  Springer.

\bibitem[{Papasavva et~al.(2021)Papasavva, Blackburn, Stringhini, Zannettou,
  and Cristofaro}]{papasavva2021qoincidence}
Papasavva, A.; Blackburn, J.; Stringhini, G.; Zannettou, S.; and Cristofaro,
  E.~D. 2021.
\newblock “Is it a qoincidence?”: An exploratory study of QAnon on Voat.
\newblock In \emph{Proceedings of the Web Conference 2021}, 460--471.

\bibitem[{Papasavva et~al.(2020)Papasavva, Zannettou, De~Cristofaro,
  Stringhini, and Blackburn}]{papasavva2020raiders}
Papasavva, A.; Zannettou, S.; De~Cristofaro, E.; Stringhini, G.; and Blackburn,
  J. 2020.
\newblock Raiders of the lost kek: 3.5 years of augmented 4chan posts from the
  politically incorrect board.
\newblock In \emph{Proceedings of the International AAAI Conference on Web and
  Social Media}, volume~14, 885--894.

\bibitem[{Raman et~al.(2019)Raman, Joglekar, Cristofaro, Sastry, and
  Tyson}]{raman2019challenges}
Raman, A.; Joglekar, S.; Cristofaro, E.~D.; Sastry, N.; and Tyson, G. 2019.
\newblock Challenges in the decentralised web: The mastodon case.
\newblock In \emph{Proceedings of the internet measurement conference},
  217--229.

\bibitem[{Ribeiro et~al.(2018)Ribeiro, Calais, Santos, Almeida, and
  Meira~Jr}]{ribeiro2018characterizing}
Ribeiro, M.~H.; Calais, P.~H.; Santos, Y.~A.; Almeida, V.~A.; and Meira~Jr, W.
  2018.
\newblock Characterizing and detecting hateful users on twitter.
\newblock In \emph{Twelfth international AAAI conference on web and social
  media}.

\bibitem[{Risch and Krestel(2020)}]{risch2020toxic}
Risch, J.; and Krestel, R. 2020.
\newblock Toxic comment detection in online discussions.
\newblock \emph{Deep learning-based approaches for sentiment analysis},
  85--109.

\bibitem[{Trautwein et~al.(2022)Trautwein, Raman, Tyson, Castro, Scott,
  Schubotz, Gipp, and Psaras}]{10.1145/3544216.3544232}
Trautwein, D.; Raman, A.; Tyson, G.; Castro, I.; Scott, W.; Schubotz, M.; Gipp,
  B.; and Psaras, Y. 2022.
\newblock Design and evaluation of IPFS: a storage layer for the decentralized
  web.
\newblock In \emph{Proceedings of the ACM SIGCOMM 2022 Conference}, SIGCOMM
  '22, 739–752. New York, NY, USA: Association for Computing Machinery.
\newblock ISBN 9781450394208.

\bibitem[{Varian(2019)}]{varian2019recent}
Varian, H.~R. 2019.
\newblock Recent Trends in Concentration, Competition, and Entry.
\newblock \emph{Antitrust Law Journal}, 82(3): 807--834.

\bibitem[{W3C(2018)}]{activitypub2018}
W3C. 2018.
\newblock ActivityPub W3C Recommendation.
\newblock \url{https://www.w3.org/TR/activitypub/}.
\newblock Accessed: 2024-04-08.

\bibitem[{Waseem and Hovy(2016)}]{waseem2016hateful}
Waseem, Z.; and Hovy, D. 2016.
\newblock Hateful symbols or hateful people? predictive features for hate
  speech detection on twitter.
\newblock In \emph{Proceedings of the NAACL student research workshop}, 88--93.

\bibitem[{Wu et~al.(2022)Wu, Williams, Simpson, and
  Semaan}]{wu2022conversations}
Wu, Q.; Williams, L.~K.; Simpson, E.; and Semaan, B. 2022.
\newblock Conversations About Crime: Re-Enforcing and Fighting Against
  Platformed Racism on Reddit.
\newblock \emph{Proceedings of the ACM on Human-Computer Interaction},
  6(CSCW1): 1--38.

\bibitem[{Yin et~al.(2023)Yin, Agarwal, Jiang, Zubiaga, and
  Sastry}]{yin2023annobert}
Yin, W.; Agarwal, V.; Jiang, A.; Zubiaga, A.; and Sastry, N. 2023.
\newblock AnnoBERT: effectively representing multiple annotators’ label
  choices to improve hate speech detection.
\newblock In \emph{Proceedings of the International AAAI Conference on Web and
  Social Media}, volume~17, 902--913.

\bibitem[{Zignani et~al.(2019)Zignani, Quadri, Galdeman, Gaito, and
  Rossi}]{zignani2019mastodon}
Zignani, M.; Quadri, C.; Galdeman, A.; Gaito, S.; and Rossi, G.~P. 2019.
\newblock Mastodon content warnings: Inappropriate contents in a microblogging
  platform.
\newblock In \emph{Proceedings of the International AAAI Conference on Web and
  Social Media}, volume~13, 639--645.

\end{thebibliography}


\section{Paper Checklist}

\begin{enumerate}

\item For most authors...
\begin{enumerate}
    \item Would answering this research question advance science without violating social contracts, such as violating privacy norms, perpetuating unfair profiling, exacerbating the socio-economic divide, or implying disrespect to societies or cultures?
    \answerYes{Yes, we have discussed about the ethical considerations in Sections 4 and 6.}
  \item Do your main claims in the abstract and introduction accurately reflect the paper's contributions and scope?
    \answerYes{Yes, we have described the main findings and contributions of this paper in the abstract and introduction.}
   \item Do you clarify how the proposed methodological approach is appropriate for the claims made? 
    \answerYes{Yes, we have discussed it in in Section~\ref{sec:methodology}.}
   \item Do you clarify what are possible artifacts in the data used, given population-specific distributions?
    \answerYes{Yes, we have discussed it in Section~\ref{sec:methodology}.}
  \item Did you describe the limitations of your work?
    \answerYes{Yes, we have described in Sections 6.3 and 7.}
  \item Did you discuss any potential negative societal impacts of your work?
    \answerYes{Yes, we have discussed in Section 6.}
      \item Did you discuss any potential misuse of your work?
    \answerYes{Yes, please check Section 6.}
    \item Did you describe steps taken to prevent or mitigate potential negative outcomes of the research, such as data and model documentation, data anonymisation, responsible release, access control, and the reproducibility of findings?
    \answerYes{Yes, we have discussed in Section 4 for data and others in Section 6.}
  \item Have you read the ethics review guidelines and ensured that your paper conforms to them?
    \answerYes{Yes}
\end{enumerate}

\item Additionally, if your study involves hypotheses testing...
\begin{enumerate}
  \item Did you clearly state the assumptions underlying all theoretical results?
    \answerNA{NA}
  \item Have you provided justifications for all theoretical results?
    \answerNA{NA}
  \item Did you discuss competing hypotheses or theories that might challenge or complement your theoretical results?
    \answerNA{NA}
  \item Have you considered alternative mechanisms or explanations that might account for the same outcomes observed in your study?
    \answerNA{NA}
  \item Did you address potential biases or limitations in your theoretical framework?
    \answerNA{NA}
  \item Have you related your theoretical results to the existing literature in social science?
    \answerNA{NA}
  \item Did you discuss the implications of your theoretical results for policy, practice, or further research in the social science domain?
    \answerNA{NA}
\end{enumerate}

\item Additionally, if you are including theoretical proofs...
\begin{enumerate}
  \item Did you state the full set of assumptions of all theoretical results?
    \answerNA{NA}
	\item Did you include complete proofs of all theoretical results?
    \answerNA{NA}
\end{enumerate}

\item Additionally, if you ran machine learning experiments...
\begin{enumerate}
  \item Did you include the code, data, and instructions needed to reproduce the main experimental results (either in the supplemental material or as a URL)?
    \answerYes{Yes, we have added all the relevant details for reproducibility or further research. We will make our code publicly available upon the paper's acceptance. We do not intend to make the data available due to privacy reasons.}
  \item Did you specify all the training details (e.g., data splits, hyperparameters, how they were chosen)?
    \answerYes{Yes, in Section 6.}
     \item Did you report error bars (e.g., with respect to the random seed after running experiments multiple times)?
    \answerYes{Yes, we report the average results and standard deviation in Section 6.2.}
	\item Did you include the total amount of compute and the type of resources used (e.g., type of GPUs, internal cluster, or cloud provider)?
    \answerYes{Yes in Section 6.1.}
     \item Do you justify how the proposed evaluation is sufficient and appropriate to the claims made? 
    \answerYes{Yes, we have discussed this in Section 4.}
     \item Do you discuss what is ``the cost`` of misclassification and fault (in)tolerance?
    \answerYes{Yes, we have discussed this in results in Section 6.}
  
\end{enumerate}

\item Additionally, if you are using existing assets (e.g., code, data, models) or curating/releasing new assets...
\begin{enumerate}
  \item If your work uses existing assets, did you cite the creators?
    \answerYes{Yes, we have cited the paper proposing the dataset.}
  \item Did you mention the license of the assets?
    \answerNA{NA}
  \item Did you include any new assets in the supplemental material or as a URL?
    \answerNA{NA}
  \item Did you discuss whether and how consent was obtained from people whose data you're using/curating?
    \answerYes{Yes, we have discussed it in Section 6.}
  \item Did you discuss whether the data you are using/curating contains personally identifiable information or offensive content?
    \answerYes{Yes, we have discussed it in Section 4.}
\item If you are curating or releasing new datasets, did you discuss how you intend to make your datasets FAIR (see \citet{fair})?
\answerNA{NA}
\item If you are curating or releasing new datasets, did you create a Datasheet for the Dataset (see \citet{gebru2021datasheets})? 
\answerNA{NA}
\end{enumerate}

\item Additionally, if you used crowdsourcing or conducted research with human subjects...
\begin{enumerate}
  \item Did you include the full text of instructions given to participants and screenshots?
    \answerNA{NA}
  \item Did you describe any potential participant risks, with mentions of Institutional Review Board (IRB) approvals?
    \answerNA{NA}
  \item Did you include the estimated hourly wage paid to participants and the total amount spent on participant compensation?
    \answerNA{NA}
   \item Did you discuss how data is stored, shared, and deidentified?
   \answerNA{NA}
\end{enumerate}

\end{enumerate}

\end{document}